\def \SAIT #1 #2 {{\em Mem.\ Soc.\ Astron.\ It.\/} {\bf #1}, #2}
\def \MESS #1 #2 {{\em The Messenger\/} {\bf #1}, #2}
\def \ASTRNACH #1 #2 {{\em Astron. Nach.\/} {\bf #1}, #2}
\def \AAP #1 #2 {{\em Astron. Astrophys.\/} {\bf #1}, #2}
\def \AAL #1 #2 {{\em Astron. Astrophys. Lett.\/} {\bf #1}, L#2}
\def \AAR #1 #2 {{\em Astron. Astrophys. Rev.\/} {\bf #1}, #2}
\def \AAS #1 #2 {{\em Astron. Astrophys. Suppl. Ser.\/} {\bf #1}, #2}
\def \AJ #1 #2 {{\em Astron. J.\/} {\bf #1}, #2}
\def \ANNREV #1 #2 {{\em Ann. Rev. Astron. Astrophys.\/} {\bf #1}, #2}
\def \APJ #1 #2 {{\em Astrophys. J.\/} {\bf #1}, #2}
\def \APJL #1 #2 {{\em Astrophys. J. Lett.\/} {\bf #1}, L#2}
\def \APJS #1 #2 {{\em Astrophys. J. Suppl.\/} {\bf #1}, #2}
\def \APSS #1 #2 {{\em Astrophys. Space Sci.\/} {\bf #1}, #2}
\def \ASR #1 #2 {{\em Adv. Space Res.\/} {\bf #1}, #2}
\def \BAIC #1 #2 {{\em Bull. Astron. Inst. Czechosl.\/} {\bf #1}, #2}
\def \JSQRT #1 #2 {{\em J. Quant. Spectrosc. Radiat. Transfer\/} {\bf #1}, #2}
\def \MN #1 #2 {{\em Mon. Not. R. Astr. Soc.\/} {\bf #1}, #2}
\def \MEM #1 #2 {{\em Mem. R. Astr. Soc.\/} {\bf #1}, #2}
\def \PLR #1 #2 {{\em Phys. Lett. Rev.\/} {\bf #1}, #2}
\def \PASJ #1 #2 {{\em Publ. Astron. Soc. Japan\/} {\bf #1}, #2}
\def \PASP #1 #2 {{\em Publ. Astr. Soc. Pacific\/} {\bf #1}, #2}
\def \NAT #1 #2 {{\em Nature\/} {\bf #1}, #2}

\documentstyle{memsait}
\input epsf.sty
\begin{opening}
\title{Recalibration of Luminosity--HI Linewidth Relations and H$_0$}
\author{R. Brent TULLY}
\institute{Institute for Astronomy\\ University of Hawaii\\
 Honolulu, Hawaii, USA}
\date{}
\end{opening}

\begin{document}

\oddpagefooter{}{}{} 
\evenpagefooter{}{}{} 
\ 
\bigskip

\begin{abstract}
A recalibration of the luminosity--linewidth technique is discussed 
which introduces (i) new cluster calibration data, (ii) new corrections for
reddening as a function of inclination, and (iii) a new zero-point calibration
using 13 galaxies with distances determined via the cepheid period--luminosity
method.  It is found that H$_0=82\pm16$~km~s$^{-1}$~Mpc$^{-1}$
(95\% confidence).

\end{abstract}

\section{Introduction}
Good relative distances to galaxies can be found from the correlations 
between their global luminosities and rotation velocities (Tully \& 
Fisher 1977).  It is true that today there are several methods with smaller 
apparent scatter, such as exploitation of the planetary nebula luminosity 
function 
(Jacoby, Ciardullo, \& Ford 1990), surface brightness fluctuations (Tonry, 
Ajhar \& Luppino 1990), and type Ia supernova peak brightnesses (Reiss, 
Press, \& Kirshner 
1996).  However the luminosity--rotation rate technique is one that can 
be applied
to a large fraction of disk galaxies, hence thousands of cases all
over the sky and out to substantial redshifts.  It remains the most
important distance tool for studies of {\it deviations} from the
universal expansion.  It can give an estimate of the Hubble Constant
with good $\sqrt N$ statistics if given a zero-point calibration. To 
make the calibration, we look at galaxies 
that obey the
correlations and which have independently known distances.

There has been recent progress in the determination of distances to 
potential calibrators which
makes it a good time to redefine the properties of the 
luminosity--linewidth correlations.  The Hubble Space Telescope (HST)
has been used to discover cepheid variable stars in 8 suitable
galaxies and to determine distances based on the period--luminosity 
relations for those stars.  Today there are 13 appropriate galaxies with 
distances determined with this method.

In addition to the new zero-point calibration there are two other
significant improvements to the methodology.  First, more high 
quality data and more complete samples are available.  Second, 
multicolor information extending to the infrared now provides better
information about obscuration to luminosities as a function of 
inclination.

\section{Three Calibration Clusters}
Malmquist bias caused by a magnitude limit (Malmquist 1920) can be
neutralized if one has a suitable calibration sample.  There are three
requirements: one, the sample should be {\it complete} to an absolute
magnitude limit fainter than the targets to be subsequently
studied; two, the intrinsic properties of the calibrators should be
{\it statistically similar} to the targets; three, the {\it quality}
of the
observations should be similar.  With sufficient effort the completeness
requirement can be met, and with sufficient care the quality of the
observed material should be comparable between data sets.  It is
impossible to be certain that the intrinsic properties of the calibrators
are truly representative but there can be some confidence in this
proposition if several alternative calibration samples, representative of
different environments, have consistent correlation slope and scatter
characteristics.

The current calibration uses 89 galaxies in three rather different clusters.
The obvious advantage of clusters is the possibility to achieve completeness
within a volume to an absolute magnitude limit.  The disadvantage is that
cluster galaxies may be systematically different from those in the field.
Two of the `clusters' used in this study are poor specimens as clusters
and were chosen because the environments are arguably the same as the
`field'.  The Ursa Major Cluster contains 62 galaxies with
$M_B^{b,i} < -16.5^m$, {\it no} ellipticals, only 10 S0-S0/$a$ types, and the
rest are spirals and irregulars with normal HI content.  The cluster
has no central core, the radial velocity dispersion is only $\sim 150$~km/s
so a crossing time is $\sim 0.5 {\rm H}_0^{-1}$, and it can be inferred
that the region is in an early stage of collapse (Tully et al. 1996).
Excluding galaxies with $i<45^{\circ}$, ${\rm T}<{\rm S}a$, and
interacting/peculiar, leaves a sample of 37 objects.
The Pisces `cluster' has been used frequently for distance scale studies
(Aaronson et al. 1986; Han \& Mould 1992) but it, too, is composed mainly of
HI-rich galaxies scattered about several insubstantial knots of early
systems and may more properly be viewed as a prominent section of an
intracluster filament (Sakai, Giovanelli, \& Wegner 1994).
Excluding galaxies with $i<60^{\circ}$, ${\rm T}<{\rm S}a$, interacting,
or with insufficient $S/N$ in the HI signal leaves a sample of 25 objects
with $M_B^{b,i} < -19^m$.
The third sample comes from the Coma Cluster, certainly a very rich and
dense environment quite unlike the field.  A calibration based only on
this cluster would be subject to justifiable criticism.  However, it is
found that  the properties of the correlations are compatible between
this extreme environment and the others which is an argument that the
methodology works over a wide range of environmental conditions.
The same exclusions as with Pisces
leaves a sample of 27 objects
with $M_B^{b,i} < -19.3^m$.
In sum, 89 galaxies are used in the present analysis that satisfy magnitude
completeness criteria.  About 20 galaxies are missing from a truly
complete sample to the present limits because the HI detections currently
have insufficient $S/N$.

\section{Revised Reddening Corrections}
Photometry in four bands from $B$ to $K^{\prime}$, available for the UMa
and Pisces
samples, make it possible to refine the corrections that must be made for
obscuration as a function of inclination.  The most sensitive tests use
deviations from mean color-magnitude
correlations involving $K^{\prime}$.  More inclined galaxies tend to be redder
because of obscuration.  Deviations from mean luminosity--linewidth
correlations can be used for independent but somewhat less sensitive tests.
More inclined galaxies tend to be fainter.

\begin{figure}
\epsfysize=13cm 
\hspace{3.5cm}\epsfbox{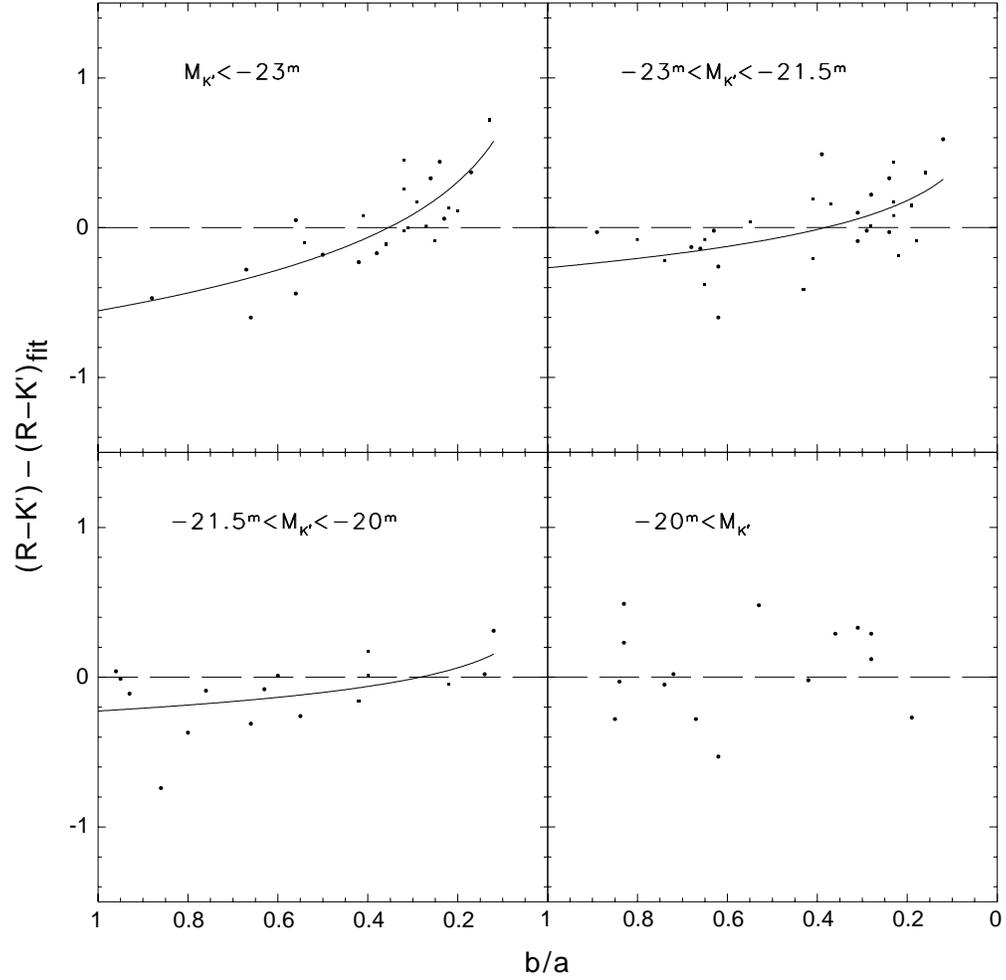} 
\caption[h]{Deviations from the mean $R-K^{\prime}$ color--magnitude
relation as a function of axial ratio.  A face-on galaxy has $b/a=1$
and an edge-on galaxy has $b/a \sim 0.2$.  The 4 panels illustrate 4
separate bins in $K^{\prime}$ luminosity, from high luminosity at the
upper left to low luminosity at the lower right.  Best fit curves are
of the form $A_R=\gamma_R{\rm log}(a/b)$.  Most of the evident effect
of reddening is in the $R$ band rather than at $K^{\prime}$.
}
\end{figure}

Reddening may be different for different classes of galaxies.
Giovanelli et al. (1995) have presented a strong case for a luminosity
dependence among spiral systems.  There is enough information now
available to specify the gross characteristics of this dependency.  The
results from the 4-band color-magnitude analysis confirms the Giovanelli
et al. claim.  The four panels of Figure~1 illustrate both the amplitude
of the reddening dependency on inclination and the dependency on
luminosity.  The plots show deviations from mean color--magnitude
relations as a function of observed axial ratio, which maps to 
inclination.  Here, $R-K^{\prime}$ colors are used.  Separate luminosity
intervals are considered in the separate panels.  There is a clear color
dependency with axial ratio for the high luminosity systems.  The
dependency diminishes to negligible for the faintest systems.  Curves 
that describe the relations
are of the form $A_R=\gamma_R{\rm log}(a/b)$.  The dependence of the
amplitude parameter $\gamma_R$ on luminosity is displayed in Figure~2.

\begin{figure}
\epsfysize=12cm 
\hspace{2cm}\epsfbox{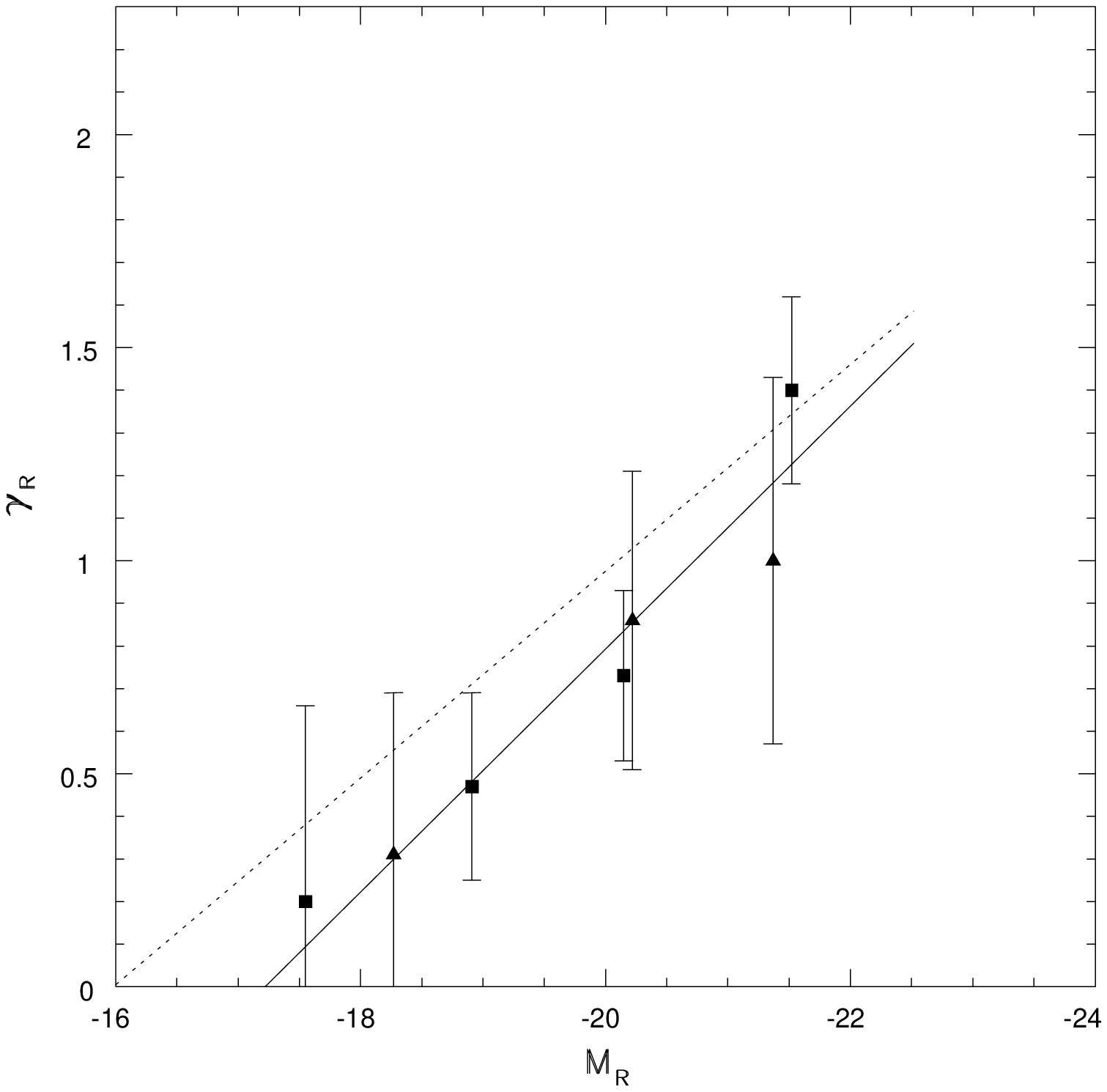} 
\caption[h]{Luminosity dependence of the obscuration amplitude parameter
in the $R$ band. {\it Boxes:} values of $\gamma_R$ corresponding to 
the curves in Fig.~1$a-d$.  {\it Triangles:} values of $\gamma_R$
determined from deviations from the mean luminosity--linewidth relation
at $R$.  {\it Solid line:} best fit to the plotted data.  {\it Dotted
line:} adopted correction; compromise between best fit shown here and
Giovanelli et al. results.
}
\end{figure}

Giovanelli et al. (1995) found a similar dependence on luminosity
though of somewhat larger amplitude.  The dashed line in Fig.~2
represents a compromise between their results and ours (translating 
from $I$ band where the Giovanelli et al. work was done) and the
corrections that are used are based on this compromise solution.  
The enhanced corrections
toward the bright end of the luminosity--linewidth relations causes
a steepening of those relations, particularly at the shorter wavelength
bands.  As a consequence, there is a considerably weaker color dependence
of the slopes than experienced previously.  The new reddening corrections
are discussed by Tully et al. (1998).

\section{Zero-Point Calibration}
The neutralization of Malmquist bias is achieved with regressions that
accept errors in the distance independent variable (ie, the linewidths)
in a sample that is complete to a limit in the distance dependent variable
(ie, magnitudes).  At a given magnitude, galaxies on either side of the
mean  linewidth are equally accessible.  It is found that the slopes of
the regressions on linewidth are compatible between the three cluster
samples.  Hence, it seems justified to slide the separate cluster
correlations in magnitude to find the relative distance differences that
result in a common correlation.    A further constraint
is provided by the requirement that the relative distances agree between
the measurements at different passbands:  the agreement between the
$B,R,I$ filters is $0.03^m$ rms.  After the superposition of the three 
clusters by minimization of the rms scatter in linewidths we 
define the {\it slopes} of the correlations and measure
the {\it relative distances} of the three clusters.  Pisces is determined
to be $2.55^m$ farther away than UMa and Coma is determined to be
$3.33^m$ farther away than UMa.

In the regime with $M_B^{b,i}<-16.5^m$,
there are now 8 galaxies with HST cepheid measurements to add to 5
galaxies with ground-based distance determinations from cepheid
observations.  These 13 galaxies lead to luminosity--linewidth
correlations that are consistent with the combined cluster slopes with
even less scatter.  There is no contradiction to the proposition that
the calibrators are similar to the cluster galaxies in properties and
provide a reasonable zero-point determination to the magnitude scale.
The $R$ band luminosity--linewidth relation for
the combined three clusters and calibrator data sets is displayed
in Figure~3.  Relations with similar scatter exist at $I$ and
$K^{\prime}$ and, with greater scatter, at $B$.  The calibrators have not
been observed at $K^{\prime}$ so, as yet, only relative distances can be
obtained at that band.

\begin{figure}
\epsfysize=12cm 
\hspace{2cm}\epsfbox{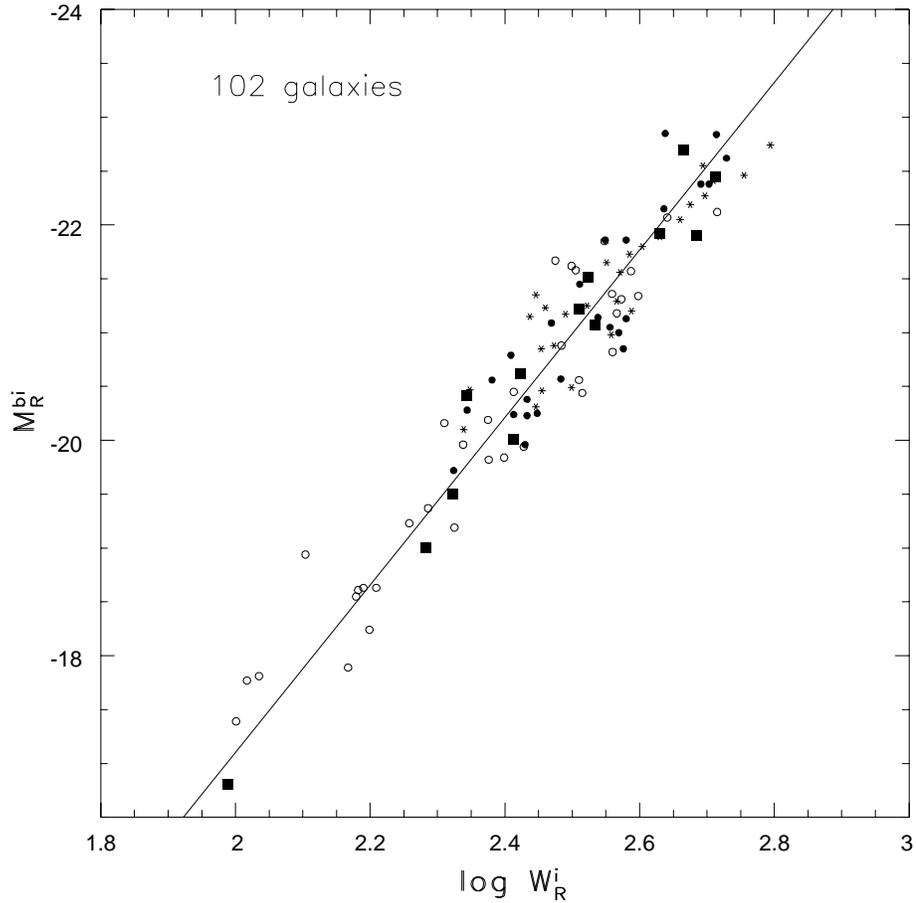} 
\caption[h]{$R$-band luminosity--linewidth relation.  The linewidth
parameter $W_R^i$ is a transformation from the observed HI 20\%
linewidth to a value statistically equal to twice the maximum rotation
velocity.  Symbols denote the various samples: {\it open circles}--UMa,
{\it filled circles}--Pisces, {\it stars}--Coma, {\it filled boxes}--
calibrators with cepheid distances.
}
\end{figure}

\section{A Determination of H$_0$}
From the combination of the $B,R,I$ fits, the Ursa Major Cluster is found 
to have a distance modulus of 31.33 corresponding to a distance of 18.5~Mpc.
From the relative distances
of the clusters, Pisces is at 60~Mpc and Coma is at 86~Mpc.
Han \& Mould (1992) report the mean redshifts of these clusters in the
microwave background frame to be 4771~km~s$^{-1}$ and 7186~km~s$^{-1}$,
respectively.  The Pisces and Coma regions are
sufficiently distant that their velocities may be a close approximation
to the Hubble expansion.  The distances to these clusters indicate
H$_0=80$ and $84$~km~s$^{-1}$~Mpc$^{-1}$, respectively.

Statistical errors are estimated to be 8\% from the typical rms scatter
of $0.40^m$ and the numbers of objects in the
samples.  There is about a 7\% uncertainty due to possible
peculiar velocities
of the two distant clusters.  The present
zero-point assumes a modulus of 18.50 for
the Large Magellanic Cloud, uncertain to 10\%.
Systematic errors are
estimated to be $<10\%$ from the observation that the scatter
in the luminosity--linewidth relations is $\sim0.4^m$ in case after case
with consistency in slopes between different environments.  Hence,
systematics, apart from occasional blunders in photometry, are probably
not a large fraction of the scatter; ie, $< 1/2 \times 0.4^m \sim 0.2$.

These results are $\sim 13\%$ higher than other recent
recalibrations of luminosity--linewidth correlations (Mould et al. 1996;
Giovanelli et al. 1997).  Lower values of H$_0$ are coming from the
distances found by the HST measurements of cepheids which
tend to give host luminosities that are brighter
at a given linewidth than found for the objects studied from the
ground.  Maybe the difference is just caused by small numbers.  There is 
about a
15\% increase of the distance scale measured in this study from the new
zero-point calibration.

On the other hand, there is a partially offsetting effect that comes about
through the new absorption corrections.  More luminous, highly inclined
galaxies are made brighter.  The galaxies in Pisces and Coma tend to be
brighter than in UMa and the samples are restricted to more inclined
systems in these clusters.  Hence, overall they receive larger corrections
and the cummulative effect is to give them smaller distances by roughly
8\%.

Compared with earlier measurements, the new cepheid calibration has
caused an increase in distances, but the revised reddening corrections
has lead to smaller distances at larger redshifts.  The overall mix
produces a lower H$_0$ than before but only by $\sim~7\%$.  It is
sobering to see these systematics of order 10\% and to appreciate that
even these matters we know about are not well resolved.
A current  estimate of the Hubble
parameter is H$_0=82\pm16$~km~s$^{-1}$~Mpc$^{-1}$ (95\% confidence)
from the
luminosity--linewidth method.  This error does not include uncertainty
in the zero-point attributable to the distance of the LMC.  There is
work yet to be done.



\end{document}